\documentclass[aip,cha,amsmath,amssymb, groupedaddress, preprint]{revtex4-1}
\usepackage{graphicx}
\usepackage[utf8]{inputenc}
\usepackage{multirow}
\usepackage{array}
\usepackage{float}
\usepackage{bm}
\usepackage{mathrsfs}
\usepackage{natbib}
\usepackage{dcolumn}
\usepackage{comment}
\usepackage{xcolor}
\usepackage{csquotes}
\usepackage{mathtools}
\usepackage[justification=justified]{caption}
\usepackage[font=small,skip=0pt]{caption}
\graphicspath{{./Figures/}}
\DeclareMathOperator*{\argmax}{arg\,max}

\begin{document}
\newcommand\reminder[1]{\textbf{\textcolor{red}{\Large{#1}}}}
\newcommand\rev[1]{\textcolor{red}{#1}}
\title{Planetary Influences on the Solar Cycle: A Nonlinear Dynamics Approach}

\author{Juan M. Muñoz}
\author{Alexandre Wagemakers}
\author{Miguel A. F. Sanjuán}
\email[]{\textbf{Author to whom correspondence should be addressed}: miguel.sanjuan@urjc.es}
\affiliation{Nonlinear Dynamics, Chaos and Complex Systems Group, Departamento de Física, Universidad Rey Juan Carlos, Tulipán s/n, 28993 Móstoles, Madrid, Spain}

\date{\today}

\begin{abstract}
We explore the effect of some simple perturbations on three nonlinear models proposed to describe large scale solar behavior via the solar dynamo theory: the Lorenz and the Rikitake systems, and a Van der Pol-Duffing oscillator. Planetary magnetic fields affecting the solar dynamo activity have been simulated by using harmonic perturbations. These perturbations introduce cycle intermittency and amplitude irregularities revealed by the frequency spectra of the nonlinear signals. Furthermore, we have found that the perturbative intensity acts as an order parameter in the correlations between the system and the external forcing. Our findings suggest a promising avenue to study the sunspot activity by using nonlinear dynamics methods. 
\end{abstract}
\maketitle

\newpage

\textbf{The Sun displays an approximately periodic activity cycle of about 11 years; this is most easily verified by looking at the number of sunspots over time. In order to explain this phenomenon, models of varying kinds have been devised, aiming to capture the main aspects of solar activity. These models focus on the dynamical relationship between electromagnetic fields and the motion of the solar plasma. This interaction is known to be complex and nonlinear. Planetary magnetic fields have been considered as a possible external factor affecting solar dynamics; we present a simple, yet novel class of models representing planetary action as perturbations of well-known dynamical models of solar activity, more specifically the Lorenz and Rikitake systems, as well as the Duffing-Van der Pol oscillator. We use these models as test beds to study how structures related to the solar cycle arise in presence of periodic planetary motions.
We find that cycle-like structures arise naturally from this approach, and we present a few aspects of the dynamics of our models that can be further explored to test how adequate they are to describe the solar cycle.}

\section{Introduction}
Over several decades, efforts in modeling various aspects of solar activity using magnetohydrodynamics (MHD) and plasma physics have been fruitful\cite{rev_2_mhdreview}, including processes in the convection zone\cite{rev_3_mhdreview_openproblems,rev_5_charbonneau} as well as coronal processes and solar wind\cite{rev_1_mhdapplied}. These nonlinear models aim to represent \textit{solar dynamo action}, which is understood as a self-sustaining interplay between the Sun's magnetic field and the fluid motion of charged material\cite{rev_4_solardynamo_rev}. Certain techniques, such as dimensional truncation and mean field treatments\cite{rev_6_meanfield}, have led to low-order dynamo models for the Sun, which allow for further analytical development\cite{stoch_dynamo,rev_14_loworder} and are computationally more viable. In particular, the $\alpha\Omega$ family of models has attracted much interest\cite{rev_7_alphaomega1} and continues to do so\cite{rev_8_alphaomega2}. This class of models provides unique visual insight into the mechanisms by which poloidal and toroidal components interact with each other. Such representations have allowed researchers to grasp the essence of dynamo action using systems of nonlinear ordinary differential equations featuring the components of the magnetic field. They typically exhibit chaotic behavior\cite{lorenz_rikitake,hanslmeier}, which is consistent with observations in solar activity.

The aim of dynamo models is to provide a framework for developing methods to explore several metrics regarding solar activity and space weather which have been defined and observed for several decades\cite{rev_9_sunspot_history}. The most prominent metric is the Wolf number, which provides a regularized quantification of the number of sunspots\cite{jaramillo1} at any given time. However, attempts so far to explain the 11-year periodicity (known as the \textit{Schwabe} cycle) in the number and motion of sunspots have been inconclusive but are still an active topic\cite{rev_10_dynamo_cycles}.

Many attempts to modify solar dynamo models to account for sunspot cycles have been endogenous in nature, e.g.~by introducing a Kuramoto model of coupled oscillators to assess anti-synchronization between northern and southern hemispheres\cite{rev_13_kuramoto}, or by adding the stochastic noise as a way of reintroducing small-scale turbulence and irregularities into the system\cite{stoch_dynamo}. We will, however, focus on an exogenous, deterministic factor that has been considered for a while and is now experiencing a revival\cite{planetary_review,ferrizmas1}: the idea that the solar dynamo may be forced externally by planetary motion. Indeed, there are several indications that coupling between the Solar magnetic field and those of nearby planets can be of significant scale\cite{rev_11_angmomentum}. In particular, it has been argued that the angular momentum distribution of the Solar System can be affected by a Jupiter-Sun exchange via a magnetic field\cite{rev_12_angmomentum}. Recently, the effect of a harmonic perturbation representing a single planet has been explored by\cite{hansson} with promising results, showing that a periodic forcing such as planetary motion can induce cycle-like behavior in well-known nonlinear models of the solar dynamo. Moreover, a multi-planetary action has been studied for a PDE model in~\cite{shaken_stirred}. These advances suggest an encouraging line of research from this simple approach featuring deterministic, tractable factors.

Our aim here is to explore the planetary paradigm further by adding several single-planet harmonic perturbations to three dynamical systems typically associated with the solar dynamo. In our case, each component of the perturbation represents the magnetic effect of one of the solar planets. We will show that simple perturbations coupled to nonlinear low-order dynamo models are a new viable way of exploring the solar cycle.

In Section II, we introduce the three main nonlinear systems we will be exploring in this article. We will also define how we represent planetary magnetic forcings via periodic perturbations, as well as how we parametrize their intensities and frequencies. We present the quantitative results of numerically integrating the perturbed dynamo systems and how these periodic perturbations introduce cycle-like behavior as well as non-trivial features in the spectral densities of the time series in Section III. Later, in Section IV, we consider to what degree the perturbation signal synchronizes with the cycle-like behavior of the perturbed system. In Section V, we study how the perturbative intensity (interpreted as overall magnetospheric intensity of the Solar System) acts as an order parameter in the cross-correlations between the perturbation and the integrated time series. Finally, in Section VI we summarize and discuss the main findings in this work, and we point out several ways in which our model can be further developed in future work.

\section{Structure of the Models}
A class of planetary and stellar dynamo models has been shown to reduce to the Lorenz system under a suitable dipolar approximation \cite{lorenz_rikitake}. More specifically, an axisymmetric $\alpha\Omega$ dynamo model is proposed
\begin{equation}
\begin{array}{cc}
    \frac{\partial A(x,t)}{\partial t} = \eta \frac{\partial^2 A}{\partial x^2}+B[\alpha_0 \cos x + C(x,t)]\\
    \frac{\partial B(x,t)}{\partial t} = \eta \frac{\partial^2 B}{\partial x^2}+V\frac{\partial A}{\partial x}\\
    \frac{\partial C(x,t)}{\partial t} = \nu \frac{\partial^2 C}{\partial x^2}+K \cdot A\cdot B\,\\
\end{array}
\end{equation}
where $A$,$B$,$C$ are symmetric, antisymmetric, and fluctuating components of the magnetic field, respectively. A second-order truncation of the Fourier mode expansion of the dynamic variables leads to the Lorenz system, provided a purely dipolar or quadrupolar mode is considered. In nondimensional units, our perturbed Lorenz system is expressed as
\begin{equation}\label{eq:lorenz0}
\begin{array}{l}
\dot{x}=\sigma(y-x) + p(t)\\
\dot{y}=x(r-z) - y \\
\dot{z}= xy - bz,
\end{array}
\end{equation}
where the perturbation $p(t)$ is introduced additively into the first equation because the square of the $x$-component is directly associated with the dynamo number and consequently with the sunspot number (in particular, the dynamo number is proportional to $x^2$). This procedure is analogous to that shown in \cite{hansson}. A classical choice of parameter values leading to dynamo-like chaotic behavior are $\sigma=10$, $b=8/3$, and $r=28$.

Another well-established low-order dynamo model of solar activity is inspired by the Rikitake system. Originally, this system was constructed as a pair of conducting discs spinning along with a wired axis, and co-inducing magnetic forces upon one another. However, a connection to magnetic reversals in planetary and stellar dynamo action was later established \cite{lorenz_rikitake}. The connection with MHD is outlined in \cite{hanslmeier}, where the current intensity through either part of the Rikitake dynamo is linked to one component of the Solar magnetic field. A simple, nondimensional version of the Rikitake system, which is obtained by imposing rotational synchronization on the two discs, is given by
\begin{equation}\label{eq:rikitake0}
    \begin{array}{l}
    \dot{x} = - \mu x + yz + p(t)\\
    \dot{y} = - \mu y + x(z-a) \\
    \dot{z} = 1 - xy.
    \end{array}
\end{equation}
The standard parameter choice $\mu=1$, $a=5$ yields chaotic behavior, and the placement of the perturbation $p(t)$ is justified analogously to the Lorenz system.

A third solar dynamo model can be easily obtained from imposing a general condition in solar MHD theory entailing dimensional truncation~\cite{dvdp} as
\begin{equation}
    \begin{array}{l}
        \frac{dB_\phi}{dt} = c_1 B_\phi + c_2 A_p - c_3 B^{3}_\phi \\
         \frac{dA_p}{dt} = c_1 A_p + c_4 B_\phi,
    \end{array}
\end{equation}
where $B_\phi$ is the toroidal component of the solar magnetic field intensity, and $A_p$ is the vector potential originating the poloidal component of the magnetic field. By combining these two equations and reparametrising appropriately, we obtain the following equation of a Duffing-Van der Pol oscillator
\begin{equation}\label{eq:dvdp0}
    \ddot{x} = \mu (1-\xi x^2) \dot{x} - \alpha x + \beta x^3 + p(t).
\end{equation}
The variable $x$ is nondimensional and proportional to $B_\phi$, and thus again to the square root of the sunspot number, which justifies the additive inclusion of $p(t)$. In \cite{dvdp_param_fit} the parameter values $\mu=0.1645$, $\xi=0.0441$, $\alpha=0.1241$, and $\beta=0.0005$ were obtained by fitting the Duffing-Van der Pol oscillator to an average solar cycle. We will use this parametric choice, which corresponds to a double-hump potential.

Since our dimensionless variables come from magnetic intensity fields, superposition grants additivity both between individual perturbations and between the global perturbation and the system. Therefore, a rather general choice of the perturbation is
\begin{equation}\label{eq:perturb_structure}
    p(t) = \frac{1}{4}\sum^{4}_{i=1}{\varepsilon_i f_i(t)},
\end{equation}
where the parameters $\varepsilon_i$ control the perturbation intensity, and the prefactor is included for normalizing $p(t)$ to the number of components. The choice of the $f_i$ functions potentially allows for different planets to affect the solar dynamo in varying ways depending on geometric or physical considerations. For simplicity, we have chosen harmonic functions, thus
\begin{equation}\label{eq:harmonic_comps}
    f_i(t) = \cos{(\omega_i t + \phi_i)}.
\end{equation}
Note that $p(t)$ is in general anharmonic unlike in the case of a single planetary perturbation. This still allows us to tune both the intensity and periodicity of each perturbation in different ways, and tailor them specifically to the numerical scale of each of the models. In order to do this, and following the criterion outlined in \cite{hansson}, we have measured an intensity scale for each of the systems by setting $\varepsilon_i=0$ (thus $p(t)=0$). Therefore, we can define the following quantities for each model
\begin{equation}\label{eq:scale_eps_omg}
        \hat{\varepsilon} = \chi \langle \lvert x^{(n)}_{0,t} \rvert \rangle \text{,}\qquad
    \hat{\omega} = 2\pi \argmax\mathscr{F}[x_{0,t}],
\end{equation}
where $\chi > 0$ modulates the perturbation intensity, and $\langle \lvert x^{(n)}_{0,t} \rvert \rangle$ is the mean absolute value of the $n$-th numerical derivative of the integrated time series $x_{0,t}$ over time; subindex 0 refers the unperturbed systems. The order $n$ is determined by the highest derivative in the equation featuring the perturbation in each model. It is $n=1$ for the Lorenz and the Rikitake models and $n=2$ for the Van der Pol-Duffing model. We have also defined an associated time scale $\hat{\omega}$ via the numerical Fourier transform of $x_{0,t}$. The above prescriptions allow us to set the perturbative parameters $\varepsilon_i$ and $\omega_i$ in terms of their intrinsic scales for each model. Thus, we must specify how we construct the $\varepsilon_i$ and $\omega_i$ in terms of these scales. For the $\varepsilon_i$, we will explore two configurations. On the one hand, a common value to all perturbations such that $\varepsilon_i = \hat{\varepsilon}$, which we will call the \textit{common} amplitude configuration. On the other hand, a set of random values drawn from a uniform distribution between $0.8\hat{\varepsilon}$ and $1.2\hat{\varepsilon}$, which we will call the \textit{mixed} configuration. Apart from representing varying perturbative intensities from each planet, this randomization avoids potential spurious effects that may arise from rational relationships between the $\varepsilon_i$, as we have no evidence of such structures within the magnetic fields of planets in the Solar System. The $\omega_i$ are similarly drawn from a uniform distribution between $0.8\hat{\omega}$ and $1.2\hat{\omega}$. Note that it is especially important to maintain an incommensurable ratio between all the $\omega_i$ to avoid periodicities that are not present in the Solar System. Note also that, since this choice of the $\omega_i$ immediately renders the whole perturbation $p(t)$ aperiodic, the choice of $\phi_i$ becomes superfluous, so we simply set $\phi_i = 0$. Also, for a given set of values of the perturbative frequencies $\omega_i$, we define a time scale $\tau$. A suitable choice is the mean angular frequency under the distribution given by normalizing the Fourier transform of $p(t)$, written formally as
\begin{equation}\label{eq:deftau}
    \frac{2\pi}{\tau} \coloneqq \big\langle \omega \big\rangle_{\hat{\mathscr{F}}[p]},
\end{equation}
where $\hat{\mathscr{F}}[p]$ denotes the Fourier transform of $p$ normalized to its integral with respect to $\omega$ over any finite interval $\Omega$ containing the support of the uniform distribution for the $\omega_i$, namely [ $0.8\hat{\omega}$, $1.2\hat{\omega}$ ]. To obtain a useful expression for the time scale, we first normalize the Fourier transform of $p$. Given that
\begin{equation}\label{eq:fourier_deltas}
    \mathscr{F}[p](\omega) \propto \sum_{i=1}^{4}{\varepsilon_i \delta(\omega-\omega_i)},
\end{equation}
and using the integral properties of Dirac delta distributions, the normalizing factor we are looking for is
\begin{equation}
    \int_\Omega \mathscr{F}[p](\omega) d\omega \propto \sum_{i=1}^{4}{\varepsilon_i},
\end{equation}
where the implicit proportionality constant is the same as in Eq.~(\ref{eq:fourier_deltas}). Thus we can write the probability distribution over $\omega$ appearing in Eq.~(\ref{eq:deftau}) in the form
\begin{equation}
    \hat{\mathscr{F}}[p] \coloneqq \bigg( \sum_{i=1}^{4}{\varepsilon_i} \bigg)^{-1} \sum_{i=1}^{4}{\varepsilon_i \delta(\omega-\omega_i)}.
\end{equation}
Its first moment, featured in Eq.~(\ref{eq:deftau}), is then computed as
\begin{equation}
    \big\langle \omega \big\rangle_{\hat{\mathscr{F}}[p]} = \int_{\Omega}\omega \hat{\mathscr{F}}[p](\omega) d\omega.
\end{equation}
Integrating, we obtain that the right-hand side of the previous equation becomes a weighted arithmetic mean of the frequencies in the form
\begin{equation}\label{eq:tau_mixed}
    \frac{1}{\tau_{mixed}} = \frac{1}{2\pi}\bigg( \sum_{i=1}^{4}{\varepsilon_i} \bigg)^{-1} \sum_{i=1}^{4}{\varepsilon_i \omega_i}.
\end{equation}
In the case of the common configuration all the weights match ($\varepsilon_i=\hat{\varepsilon}$), thus becoming a standard arithmetic mean
\begin{equation}\label{eq:tau_common}
   \frac{1}{\tau_{common}} = \frac{1}{2\pi} \bigg(\frac{1}{4} \sum_{i=1}^{4}{\omega_i}\bigg) .
\end{equation}

Lastly for this Section, we must set an upper bound on the intensity parameter $\chi$ if we aim to represent relatively small perturbations. Specifically, we may look at the absolute value of the perturbation $\langle \lvert p(t) \rvert \rangle$ averaged over time and over realizations of the random variables $\varepsilon_i$, which we can then compare with $\langle \lvert x^{(n)}_{0,t} \rvert \rangle$ as defined above. Using the triangular inequality as well as the linear properties of the expected value, then computing the latter over the $\varepsilon_i$ under the uniform distribution described earlier (note that this last step gives the same result as using the common configuration where $\varepsilon_i=\hat{\varepsilon}$) yields

    \begin{align}
    \langle \lvert p(t) \rvert \rangle _{\varepsilon} =  \frac{1}{4} \Bigg\langle \Bigg\lvert \sum_{i=1}^{4}{ \varepsilon_i \cos(\omega_i t)} \Bigg\rvert \Bigg\rangle_{\varepsilon} \leq \frac{1}{4} \Bigg\langle  \sum_{i=1}^{4}{\varepsilon_i \big\lvert \cos(\omega_i t)} \big\rvert \Bigg\rangle_{\varepsilon} = \\ \notag  = \frac{1}{4}   \sum_{i=1}^{4}{\langle \varepsilon_i\rangle_{\varepsilon} \big\lvert \cos(\omega_i t)} \big\rvert  =  \frac{\hat{\varepsilon}}{4} \sum_{i=1}^{4} {\big\lvert \cos(\omega_i t) \big\rvert}.
    \end{align}

Then, averaging over time in each absolute signal is equivalent to averaging over half a period of the signed signal, yielding
\begin{equation}
      \langle \lvert p(t) \rvert \rangle _{\varepsilon,t} \leq \frac{\hat{\varepsilon}}{4} \sum_{i=1}^{4}{\bigg\langle \big\lvert \cos(\omega_i t)} \big\rvert \bigg\rangle _{t} = \frac{\hat{\varepsilon}}{4}\sum_{i=1}^{4}{\frac{2}{\pi}} = \frac{2\hat{\varepsilon}}{\pi} ,
\end{equation}
where the upper bound becomes independent of the realizations of the $\omega_i$. We then impose that our upper bound to the mean absolute perturbation $\langle \lvert p(t) \rvert \rangle$ be less than the mean absolute derivative in the unperturbed system $\langle \lvert x^{(n)}_{0,t} \rvert \rangle$. Using the definition of $\hat{\varepsilon}$, we obtain the following bound on the intensity parameter $\chi$

\begin{equation}\label{eq:chi_bound}
         \langle \lvert p(t) \rvert \rangle \leq  \frac{2\hat{\varepsilon}}{\pi} = \frac{2\chi \langle \lvert x^{(n)}_{0,t} \rvert \rangle}{\pi} < \langle \lvert x^{(n)}_{0,t} \rvert \rangle \Leftrightarrow \chi < \frac{\pi}{2}.
\end{equation}
Therefore, $\chi < \pi/2$ is a sufficient condition to remain in the small perturbation regime in a mean-absolute sense. We will observe this requirement in the following sections.

Regarding computational resources, we have integrated all of the differential equations using stiff BDF methods from Python's NumPy \enquote{odeint} wrapper for Fortran's ODEPACK LSODA routine. The Fourier analysis on the integrated time series $x_t$ is computed using  NumPy's \enquote{fft} library.

\newpage

\begin{table}
\scalebox{0.9}{
\begin{ruledtabular}
\begin{tabular}{cccc}
                     &  Lorenz & Rikitake & Duffing-VdP   \\
\hline
$n$  & 1 & 1 & 2 \\
  $\langle|x^{(n)}_{0,t}|\rangle$  & 33.53  & 2.698    & 0.6326 \\
 $\chi$ &6/5 & 3/40 & 1/5\\
 $\hat{\omega}/2\pi~(\times 10^{-3})$  &4.80 & 2.25   & 8.62 
\end{tabular}
\end{ruledtabular}
}
\end{table}
\captionof{table}{\textbf{Parameter values of the perturbative intensity and the frequency in our models.} For each of our three dynamical systems, the baseline values (which vary by a random factor ranging from 0.8 to 1.2 in the mixed configuration) for the perturbative parameters are detailed above. The baseline amplitudes are proportional to $\langle|x^{(n)}_{0,t}|\rangle$, which is the mean absolute value of the $n$-th derivative of the time series $x_t$ obtained by numerically integrating each of our systems under no perturbation; the value of $n$ for each system is also detailed above. The amplitudes are also modulated by $\chi$, whose values we have set in order to represent relatively small, yet qualitatively meaningful perturbative intensities. Finally, the linear frequencies shown in the last row have been obtained by maximizing the numerical Fourier transform $\mathscr{F}[x_t]$ of the signal in each unperturbed system.\label{tab:params}}

\section{Integrated time series and spectral analysis}
The main parameter values that we have used are shown in Table~\ref{tab:params}, where $\langle|x^{(n)}_{0,t}|\rangle$ and $\hat{\omega}/2\pi$ have been measured numerically as defined in Eq.~(\ref{eq:scale_eps_omg}). The values of $\chi$ have been obtained after a careful exploration of the parameter space.

\begin{figure}[H]
    \centering
    \makebox[\textwidth][c]{\includegraphics[width=1.2\textwidth,height=.6\textheight]{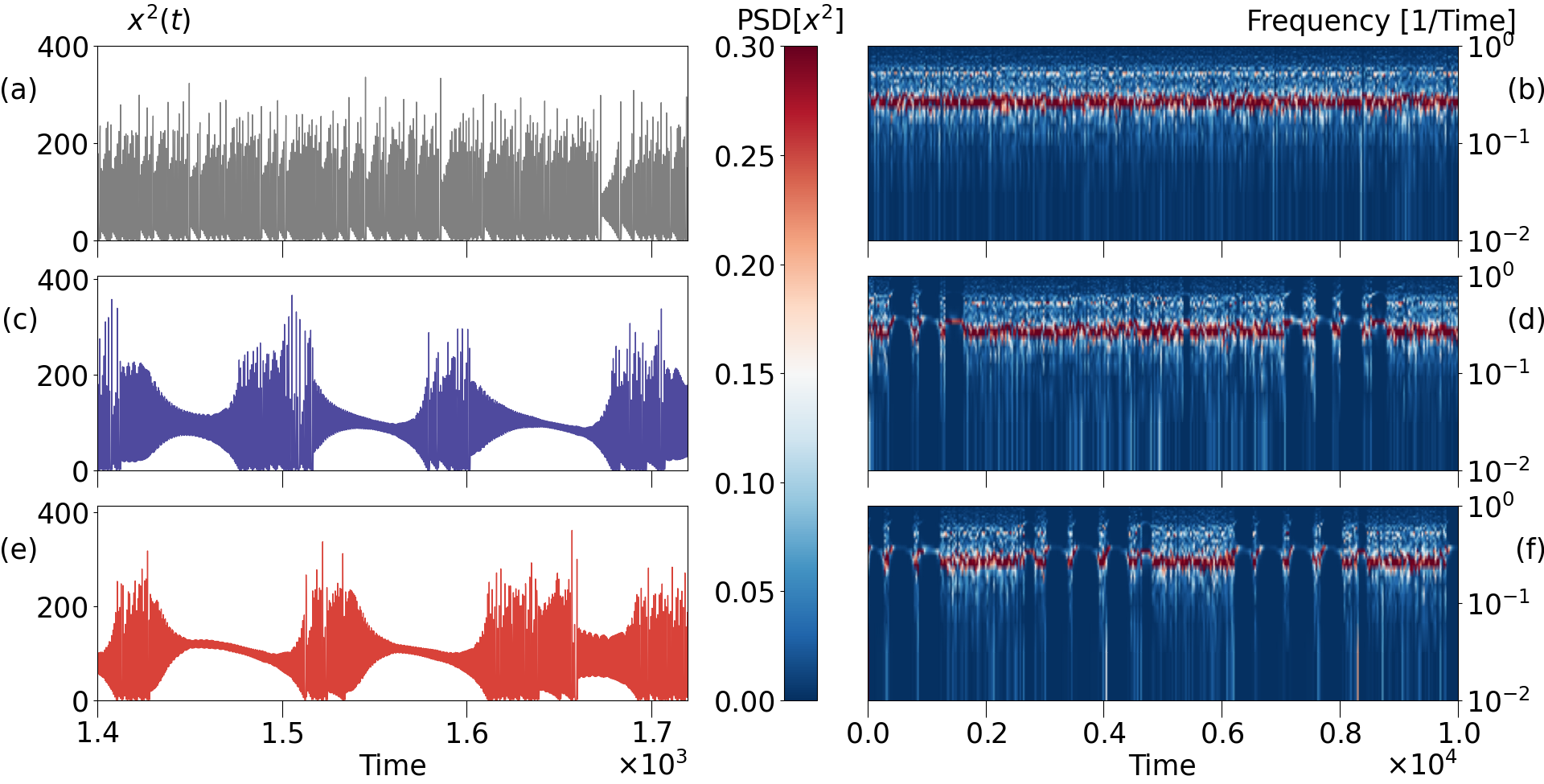}}
        \caption{\textbf{Cycle-like structures arising in the perturbed Lorenz system and associated frequency domain structures.} (a-b) Unperturbed system. (c-d) Common amplitude configuration ($\varepsilon_i=\chi\hat{\varepsilon}$). (e-f) Mixed amplitude configuration ($\varepsilon_i/\chi\hat{\varepsilon}\sim\mathcal{U}(0.8,1.2)$). The time scale in the left panels has been reduced to display the intermittent structures in $x^2$, which is proportional to the sunspot count (known as the \textit{Wolf number}) in the corresponding solar dynamo model for the Lorenz system. Spectrograms are displayed in panels (b), (d) and (f), where we show how the intrinsic structure of the spectral density in the unperturbed system becomes coupled with new frequency components when either perturbation is included. Hence, the natural frequencies become intermittent in the common and mixed configurations ((d) and (f) respectively), but more notably so in the mixed configuration.\label{fig:lorenz_spectrum}}
\end{figure}

\begin{figure}[H]
    \centering
    \makebox[\textwidth][c]{\includegraphics[width=1.2\textwidth,height=.6\textheight]{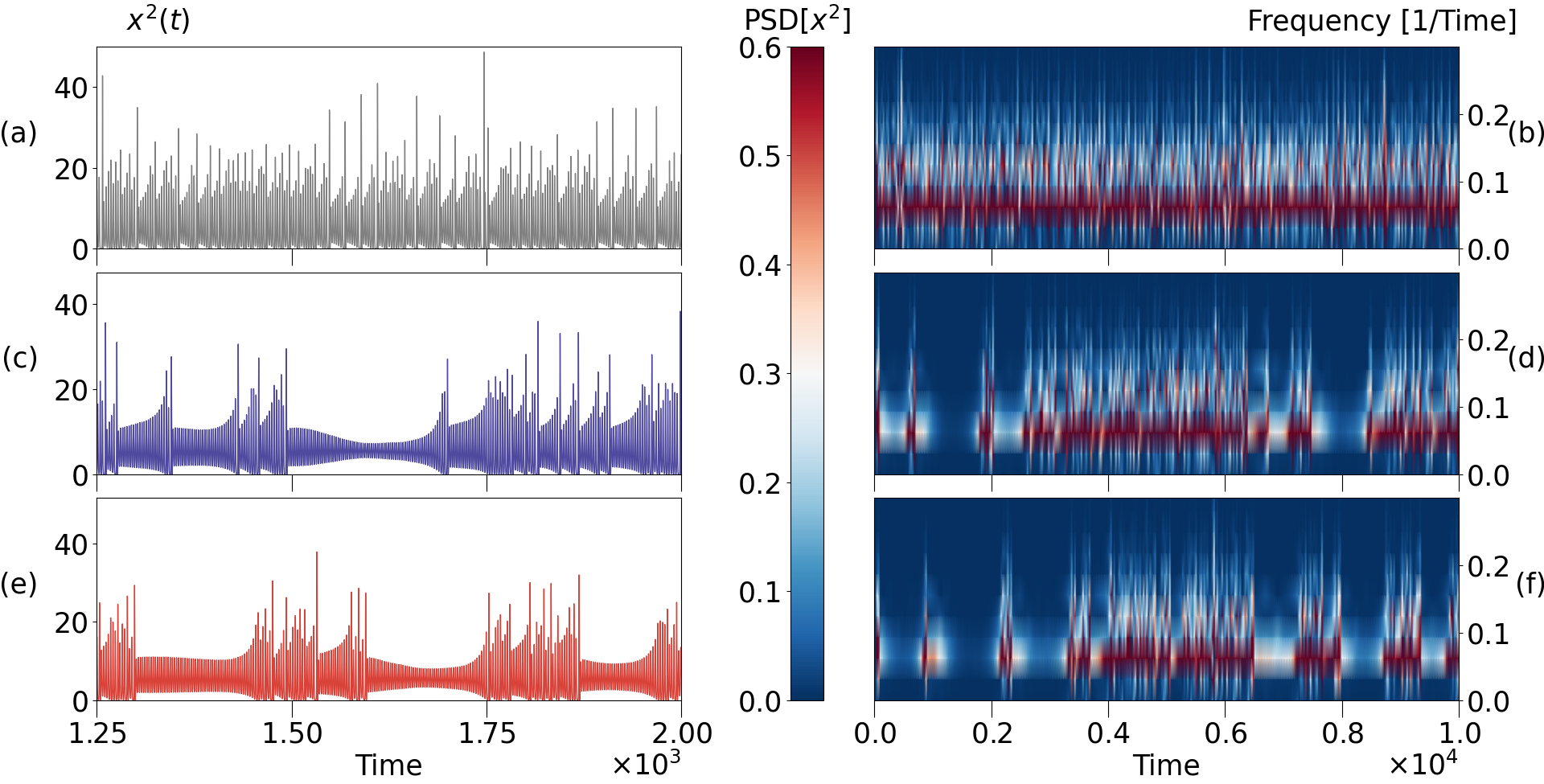}}
       \caption{\textbf{Perturbation-associated cyclic structures in the Rikitake system and spectral domain features.} (a-b) Unperturbed system. (c-d) Common amplitude configuration ($\varepsilon_i=\chi\hat{\varepsilon}$). (e-f) Mixed amplitude configuration ($\varepsilon_i/\chi\hat{\varepsilon}\sim\mathcal{U}(0.8,1.2)$). The time scale in the left panels has been reduced to display the intermittent structures in $x^2$, which is proportional to sunspot count (known as the \textit{Wolf number}) in the corresponding solar dynamo model for the Rikitake system. Spectrograms are displayed in panels (b), (d) and (f), where we show that the spectral density is not enhanced anywhere by the perturbation, but instead most of the temporal structure is suppressed at intermittent intervals.\label{fig:rikitake_spectrum}}
\end{figure}

\begin{figure}[H]
    \centering
    \makebox[\textwidth][c]{\includegraphics[width=1.2\textwidth,height=.6\textheight]{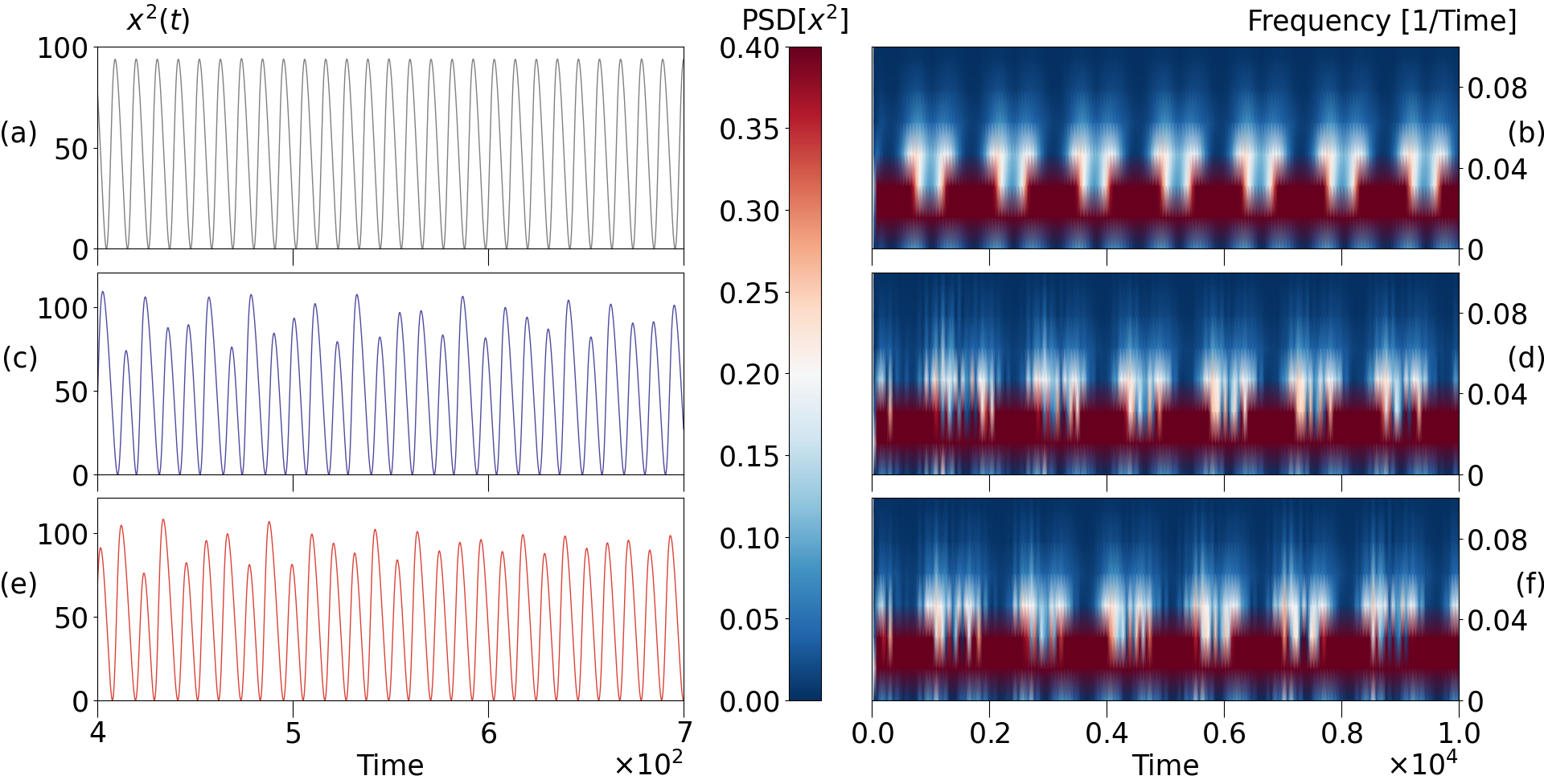}}
       \caption{\textbf{The effect of adding harmonic perturbations on the Duffing-Van der Pol oscillator with associated structures in the frequency domain.} (a-b) Unperturbed system. (c-d) Common amplitude configuration ($\varepsilon_i=\chi\hat{\varepsilon}$). (e-f) Mixed amplitude configuration ($\varepsilon_i/\chi\hat{\varepsilon}\sim\mathcal{U}(0.8,1.2)$). The time scale in the left panels has been reduced to display the aperiodic, cycle-like structures in $x^2$, which is proportional to the sunspot count (known as the \textit{Wolf number}) in the corresponding solar dynamo model for the Duffing-Van der Pol oscillator. Unlike the Lorenz and Rikitake systems, the Duffing-Van der Pol does not display intermittency, but rather shows aperiodic oscillatory behavior resembling solar cycles over yearly scales. Spectrograms are displayed in panels (b), (d) and (f), where we show that the periodic spectral density of the unperturbed oscillator gains spectral power at high frequencies that survive for shorter periods of time. \label{fig:dvdp_spectrum}}
\end{figure}

In Figs.~\ref{fig:lorenz_spectrum}-\ref{fig:dvdp_spectrum} we show the main results of this article. Here we focus on the $x^2$ variable for all of the models, since it is proportional to the sunspot number and thus a good indicator of solar activity cycles. We have also included spectrograms to explore the temporal stability of the frequency components that give rise to the oscillatory, aperiodic structures representing solar cycles.

Each model displays a combination of nonlinear, aperiodic oscillatory behavior with other temporal structures induced by the multi-planetary perturbations that are better observed in the frequency domain. The Lorenz system shows how the introduction of perturbations enhances a band of lower frequencies. They correspond to larger, cycle-like structures beyond the oscillatory behavior associated to orbits around the foci of the attractor. Another noticeable effect is the intermittency of the spectral information induced by the perturbations. This phenomenon is appreciably more noticeable in the common configuration. The Rikitake system displays a similar behavior where the spectral intermittency is intensified by the randomization of the amplitudes. The Duffing-Van der Pol oscillator is markedly more coherent in its temporal structure, due to its nonlinear yet non-chaotic nature. Here, the introduction of perturbations creates shorter scale structures in the spectrogram, showing interferential features in both perturbed configurations versus the unperturbed one. 

\newpage
\newpage
\newpage

\section{The perturbations couple with solar dynamo models}
\begin{figure}[H]
    \centering
    \makebox[\textwidth][c]{\includegraphics[width=1.2\textwidth,height=.6\textheight]{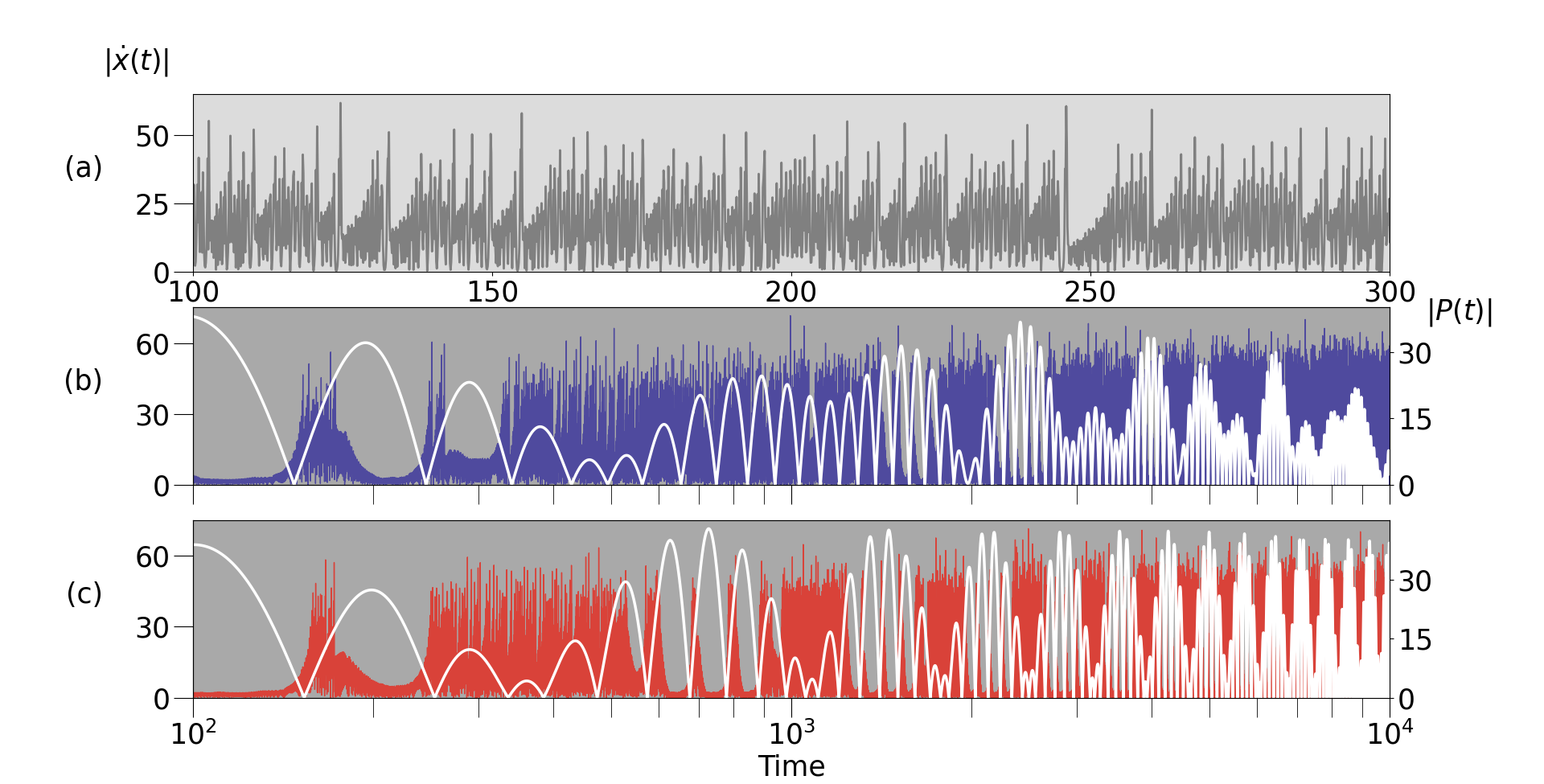}}
        \caption{\textbf{Coupling between perturbation and velocity of the $x$ component in the Lorenz system.} (a) The unperturbed system is shown over a shorter time period to show the small-scale structure of $\lvert\dot{x}(t)\lvert$. (b) Common amplitude configuration ($\varepsilon_i=\chi\hat{\varepsilon}$). (c) Mixed amplitude configuration ($\varepsilon_i/\chi\hat{\varepsilon}\sim\mathcal{U}(0.8,1.2)$). The intermittency patterns shown in the two panels on the bottom are noticeably driven by the perturbation. Thus, a direct form of coupling is shown when the additive components of the perturbation are harmonic. Non-intermittent periods where the system behaves closer to the unperturbed system coincide with a destructive interference among the components of the perturbation. \label{fig:lorenz_V-P}}
\end{figure}

\begin{figure}[H]
    \centering
    \makebox[\textwidth][c]{\includegraphics[width=1.2\textwidth,height=.6\textheight]{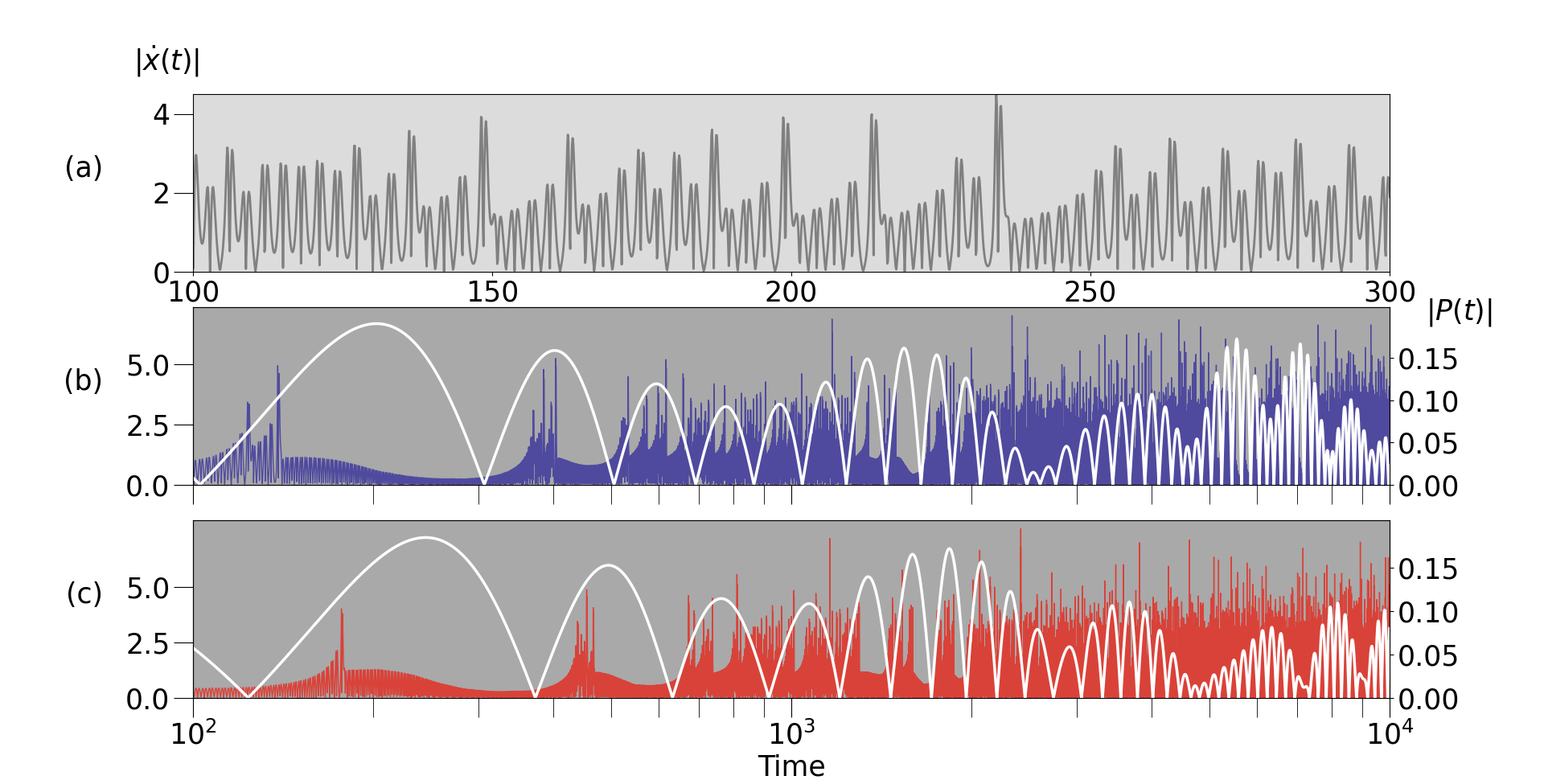}}
       \caption{\textbf{Coupling in the perturbed Rikitake system between planetary forcing and velocity of the $x$ component.} (a) The unperturbed system is shown over a shorter time period to show the small-scale structure of $\lvert\dot{x}(t)\lvert$, where we can see a double-peak structure in each cycle, and a some degree of saw-like periodicity over $4$-$6$ double cycles. (b) Common amplitude configuration ($\varepsilon_i=\chi\hat{\varepsilon}$). (c) Mixed amplitude configuration ($\varepsilon_i/\chi\hat{\varepsilon}\sim\mathcal{U}(0.8,1.2)$). In this system, the perturbative intensity is relatively weak yet the system shows clear signs of coupling under additive perturbations. The $x$ component of velocity is driven by the perturbation, even though the latter is about $2$ orders of magnitude smaller than the velocity in absolute terms.\label{fig:rikitake_V-P}}
\end{figure}

\begin{figure}[H]
    \centering
    \makebox[\textwidth][c]{\includegraphics[width=1.2\textwidth,height=.6\textheight]{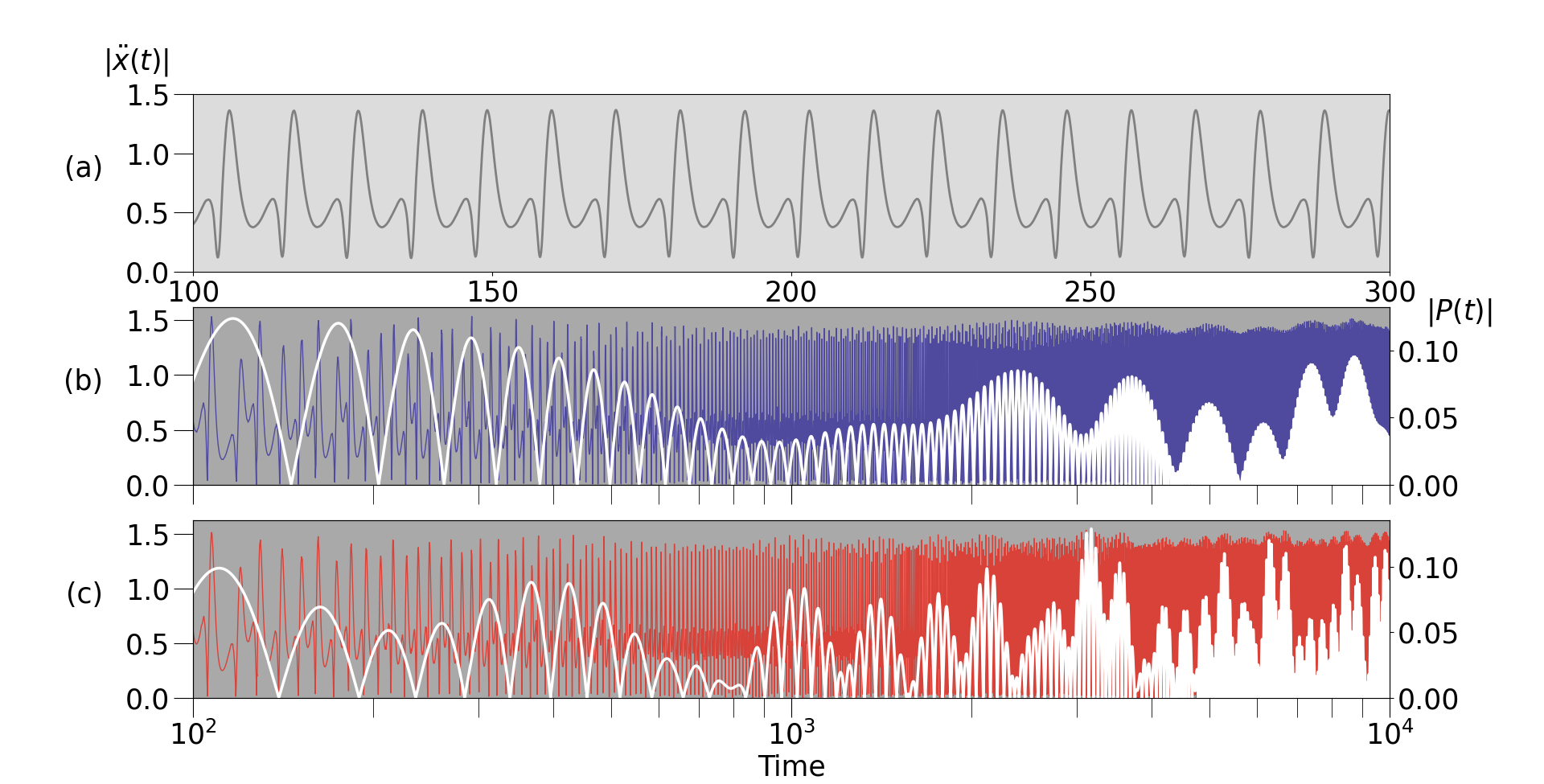}}
       \caption{\textbf{Harmonic perturbations couple indirectly with acceleration in the Duffing-Van der Pol oscillator.} (a) The unperturbed time series is shown over a shorter time period to show the small-scale structure of $\lvert\ddot{x}(t)\lvert$, where we see a periodic double pulsation. (b) Common amplitude configuration ($\varepsilon_i=\chi\hat{\varepsilon}$). (c) Mixed amplitude configuration ($\varepsilon_i/\chi\hat{\varepsilon}\sim\mathcal{U}(0.8,1.2)$). Additive harmonic perturbations do not cause intermittency in the Duffing-Van der Pol oscillator. Instead, they drive an aperiodic superposition of envelopes modulating oscillation amplitudes. This indicates that non-trivial spectral features are produced by the coupling between the system and the perturbation, even though the latter is about $2$ orders of magnitude smaller than the acceleration in absolute terms. \label{fig:dvdp_V-P}}
\end{figure}

In Figs.~\ref{fig:lorenz_V-P}-\ref{fig:dvdp_V-P} we show the absolute value of the numeric differentiate series $x^{(n)}_t$  ($n=1$ for Lorenz and Rikitake and $n=2$ for the Duffing-Van der Pol oscillator) as shown in Eqs.~(\ref{eq:lorenz0}), (\ref{eq:rikitake0}) and (\ref{eq:dvdp0}).

This perspective provides a few insights. First, irrespective of how small the perturbations are in relation to the signals, they both become effectively coupled: the relative scales range from $0.5$ to about one order of magnitude. Secondly, the behavior of the Lorenz and Rikitake systems is directly linked to the perturbations. As a matter of fact, wherever there is a destructive interference between the harmonic components of the perturbation, the systems display highly nonlinear behavior similar to their corresponding unperturbed cases. However, for periods of stronger perturbative amplitude, the systems couple more strongly with the perturbation causing the signal to appear more deterministic. The aperiodic nature of the perturbation creates the irregular intermittency between the two regimes. Overall, intermittency is stronger for common amplitudes than for mixed amplitudes both in the Lorenz and Rikitake systems. Finally, the Duffing-Van der Pol oscillator display a weaker response to the perturbations, where the oscillations in $p(t)$ originate a superposition of envelopes in the oscillations. This results in cycle-like structures with irregular amplitudes that are nonetheless deterministic.
\newpage
\newpage
\newpage
\newpage

\section{Perturbative intensity as an order parameter\label{sec:order_param}}
In order to explore how much information of the perturbation is also contained in the perturbed signal, we have studied cross-correlations in the absolute values of the time series of the perturbation $p_t$ and the time derivative of the signal to the order corresponding to each model. Thus, we define incremental variables
\begin{equation}
    \Delta p_t = |p_t| - |p_{t-1}|,\text{     }\Delta x^{(n)}_t = |x^{(n)}_t| - |x^{(n)}_{t-1}|
\end{equation}
from the absolute time series and perturbation, where the superscript $n$ indicates numerical differentiation of $x_t$ to $n$-th order. This is done to factor out a spurious component in the cross-correlation coming from auto-correlations in the non-incremental variables. Then, we define a time displacement variable $t'$ in order to quantify correlations at different delays. Thus, for every value of the perturbative intensity $\chi$, we run a simulation over $3000$ time units for each perturbative configuration (common and mixed) and then compute the cross-correlation $\rho$ as the normalised dot-product
\begin{equation}
    \rho(t',\chi) = \frac{1}{\|\Delta p\|\cdot\|\Delta x^{(n)}\|}\sum_t {\Delta p_t\cdot\Delta x^{(n)}_{t+t'} }
\end{equation}
over the chosen time span. Note that the cross-correlation is standardised in the sense that $-1\leq\rho\leq 1$. Note also that causality requires $t'\geq 0$ under the above definition.

\begin{figure}[H]
    \centering
    \makebox[\textwidth][c]{\includegraphics[width=1.2\textwidth,height=.6\textheight]{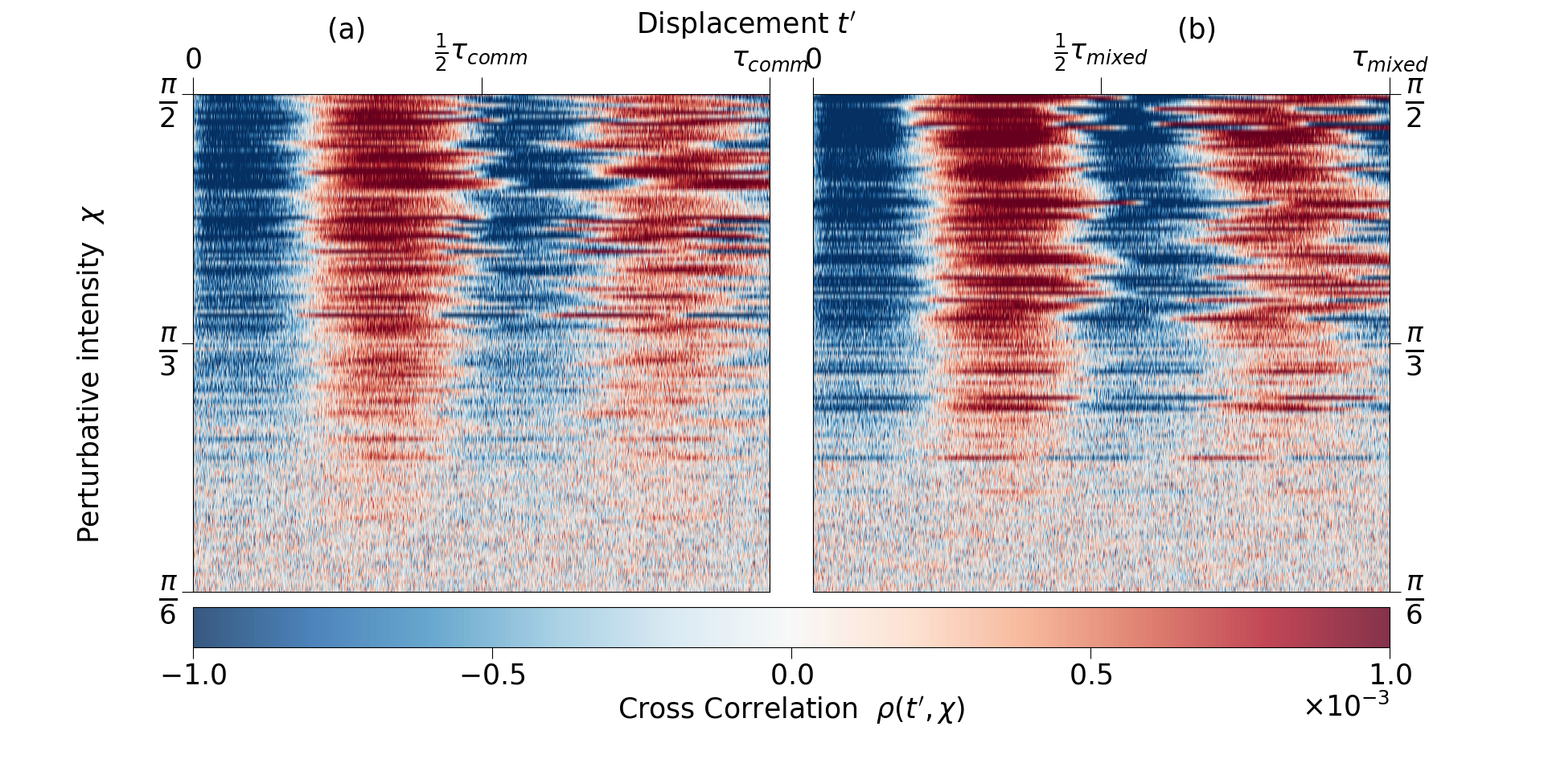}}
        \caption{\textbf{Cross correlations between the absolute perturbation $p(t)$ and the absolute velocity of the integrated time series $x_t$ in the Lorenz system.} Perturbative intensity is modulated by the $\chi$ parameter as per Eq.~(\ref{eq:scale_eps_omg}). The time displacement used in the cross-correlation ranges between $0$ and the natural time scale of the perturbation given by $\hat{\omega}$ in Eqs.~(\ref{eq:tau_mixed}-\ref{eq:tau_common}). (a) Common amplitude configuration ($\varepsilon_i=\chi\hat{\varepsilon}$). (b) mixed amplitude configuration ($\varepsilon_i/\chi\hat{\varepsilon}\sim\mathcal{U}(0.8,1.2)$). The time scale given by each corresponding $\tau$ is approximately two full periods in cross-correlation, each with half-periods of opposite sign. A smooth transition takes place throughout the range, where the cross-correlation swings with more amplitude over the time displacement from about $\pi/4$, although slightly sooner for the common configuration. Correlations are thus a relatively smooth function of the perturbative intensity in any range of $\chi$ for any time displacement within the range of our observations.\label{fig:lorenz_correl}}
\end{figure}

\begin{figure}[H]
    \centering
    \makebox[\textwidth][c]{\includegraphics[width=1.2\textwidth,height=.6\textheight]{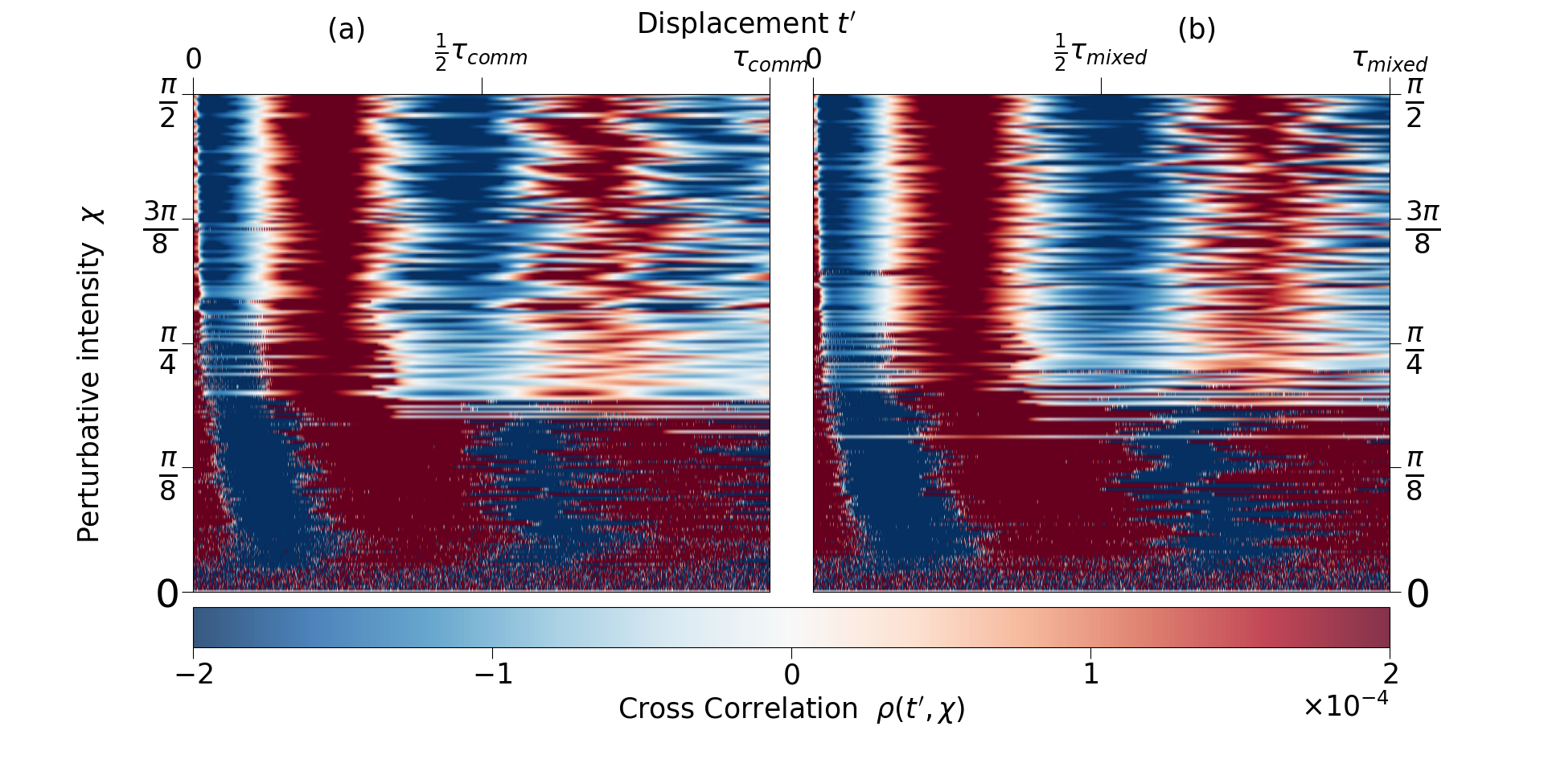}}
       \caption{\textbf{Cross correlations between the integrated time series of the absolute perturbation $\lvert p(t)\rvert$ and the absolute velocity $\dot{x}_t$ in the Rikitake system.} Eq.~(\ref{eq:scale_eps_omg}) describes how the perturbative intensity is modulated by the parameter $\chi$. We have used a time displacement variable in the cross-correlation up to the natural time scale of the perturbation given by $\hat{\omega}$ in Eqs. ~(\ref{eq:tau_mixed}-\ref{eq:tau_common}). (a) Common amplitude configuration ($\varepsilon_i=\chi\hat{\varepsilon}$). (b) mixed amplitude configuration ($\varepsilon_i/\chi\hat{\varepsilon}\sim\mathcal{U}(0.8,1.2)$). The time scale given by each corresponding $\tau$ is approximately two full periods in cross-correlation, each with half-periods of opposite sign. Two sharp transitions in the cross-correlation are observable as a function of $\chi$. First, the alternating relatively periodic regime begins at about $\chi=\pi/48$. Below this point, there is little to no structure in the cross-correlation as a function of displacement. Second, we see a transition from longer, more clear alternating periods into shorter periods transitioning more smoothly into one another. This takes place roughly at $\chi=\pi/4$.\label{fig:rikitake_correl}}
\end{figure}

\begin{figure}[H]
    \centering
    \makebox[\textwidth][c]{\includegraphics[width=1.2\textwidth,height=.6\textheight]{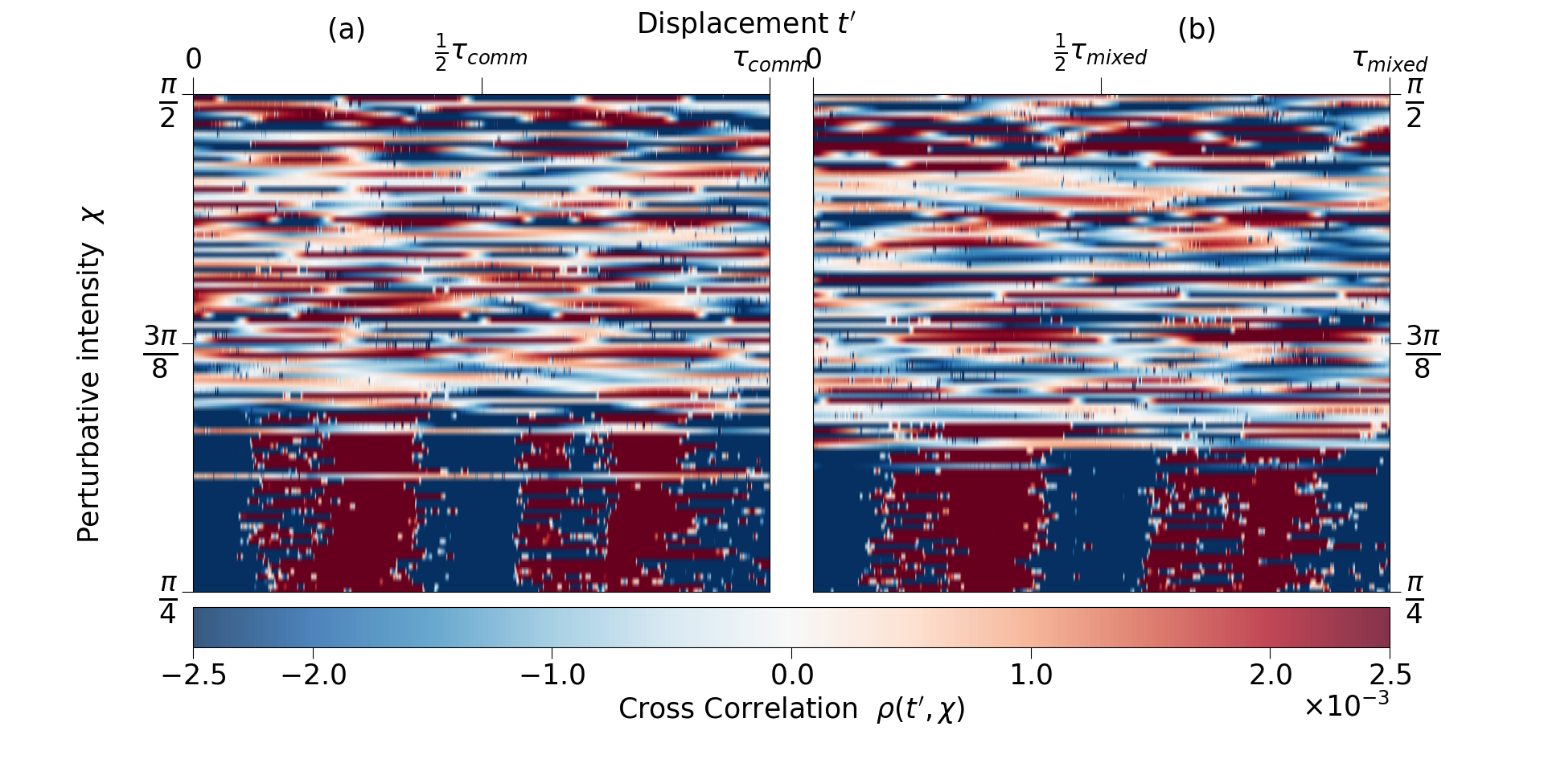}}
       \caption{\textbf{Cross correlations in the Duffing-Van der Pol oscillator for different perturbative intensities between the absolute external forcing and the absolute acceleration of the integrated time series $\ddot{x}_t$.} The $\chi$ parameter modulates the perturbative intensity as shown in Eq.~(\ref{eq:scale_eps_omg}). The time displacement variable ranges up to the natural time scale of the perturbation given by $\hat{\omega}$ in Eqs.~(\ref{eq:tau_mixed}-\ref{eq:tau_common}). (a) Common amplitude configuration ($\varepsilon_i=\chi\hat{\varepsilon}$). (b) mixed amplitude configuration ($\varepsilon_i/\chi\hat{\varepsilon}\sim\mathcal{U}(0.8,1.2)$). The time scale given by each corresponding $\tau$ is approximately two full periods in cross-correlation, each with half-periods of opposite sign. There is an intermittent transition below $\chi=3\pi/8$, where the quasi-periodic structure of the cross-correlation as a function of the time displacement $t^{\prime}$ is lost at irregular values of $\chi$. This entails that $\rho$ is no longer a smooth function of $\chi$ for perturbative intensities above the critical point. The mixed configuration shows the critical point more clearly. It also lies somewhat below that of the common configuration.\label{fig:dvdp_correl}}
\end{figure}

In Figs.~\ref{fig:lorenz_correl}-\ref{fig:dvdp_correl} we show $\rho(t',\chi)$ for our three systems in both the common and mixed configurations. The delay variable spans up to the time scale of the perturbation that we defined in Eqs. ~(\ref{eq:tau_mixed}-\ref{eq:tau_common}). The parameter $\chi$ governing perturbative intensity spans up to the limit that we established in Eq.~(\ref{eq:chi_bound}) in order to remain in the small perturbation regime.

Perturbative intensity shows itself as an order parameter in the cross-correlation between the perturbed time series and the perturbation itself for our three models, and for both common and mixed configurations, consistently over the time displacement window considered. Additionally, the time scales $\tau_{comm}$ and $\tau_{mixed}$ defined by the perturbation in both configurations are visibly related to the time scales of observable $\rho$ in an approximate proportion of $1:2$.

All three models display unique features. The Lorenz system transitions smoothly into a regime where the signal becomes increasingly correlated with the perturbation for higher intensities. The Rikitake system shows two critical points in $\chi$: from uncorrelated to sharply correlated, then from sharply correlated to loosely correlated. The Duffing-Van der Pol oscillator shows a relatively sharp transition. At low $\chi$, there is an ordered phase where correlation switches from positive to negative correlation by changing $t^{\prime}$ and more or less independently of $\chi$. However, at higher values of $\chi$, we find a disordered phase where correlation oscillates less sharply with $t^{\prime}$ and, more interestingly, very sensitively with $\chi$. In fact, for some values of the perturbative intensity we see almost no correlation at all (visible as thin white horizontal stripes). The critical point lies somewhat below $3\pi/8$, with the mixed configuration showing a sharper transition.

\newpage
\newpage
\newpage
\newpage

\section{Discussion and conclusions}
Looking toward solar magnetohydrodynamics, the relatively periodic nature of sunspot cycles and magnetic inversion has remained a prominent feature of the system that has found no satisfactory answer. On the other hand, Jupiter's orbital cycle is about $11.86$ years, which is not far from the roughly-$11$-year Schwabe cycle. We also know that Jupiter's mass is remarkably comparable to that of the Sun, at about $1\%$ of the latter. Indeed, it has been argued \cite{ang_momentum} that Jupiter may be carrying a large portion of the Solar System's angular momentum due to magnetic coupling with the Sun, as opposed to the predictions of standard models of the formation of the Solar System. It is well documented that small perturbations, such as this tidal-like proposed mechanism, can profoundly affect the behavior of chaotic nonlinear systems, which the solar dynamo has been extensively argued to be. It has also been argued that the joint action of the four most influential planets in magnetic terms, both by distance and magnetospheric intensity (Venus, Earth, Mars and Jupiter) may be a better approximation to magnetic tidal effects upon the Sun's MHD system.

We have observed cycle-like structures in our perturbed models. They display varying amplitudes and relatively periodic behavior (left and right panels in Figs.~\ref{fig:lorenz_spectrum}-\ref{fig:dvdp_spectrum}, respectively). Both perturbative configurations are qualitatively compatible with our observations of the Schwabe cycle.

Moreover, the different phenomenologies arising in either of the three underlying systems (Lorenz, Rikitake and Duffing-Van der Pol) allow us to discuss their relative merits when our perturbation is included.

The Lorenz system shows most clearly how introducing harmonic perturbative signals creates a cycle-like pattern beyond the high-frequency oscillations shown in the unperturbed case (similarly to how the solar cycle modulates daily fluctuations in the Wolf number). Also, the coupling between the perturbation and the first derivative of the $x$-component (proportional to dynamo action and thus to the Wolf number) suggests a new methodological approach to studying the effects of magnetic coupling between solar MHD and the planets. The Rikitake system suggests an interesting possibility for the solar cycle. Rather than the external planetary perturbations originating the cycle directly, it would be originating from solar dynamics itself when some frequency components are suppressed by the perturbation (the spectrograms in Fig.~\ref{fig:rikitake_spectrum} show this very clearly). The coupling between the perturbation and the $x$-component, as for the Lorenz system, also allows for experimental testing of the modeling hypotheses we have worked upon. Moreover, this suggests new ways of assessing the comparative merits of both solar dynamo models from an observational perspective. The Duffing-Van der Pol oscillator shows an aperiodic oscillatory structure whose large-scale patterns (i.e. over time periods larger than the oscillations) do not mimic the structure of the perturbation (see Fig.~\ref{fig:dvdp_V-P}). In this sense, the succint effect of the planetary perturbation on the Duffing-Van der Pol oscillator is more representative of solar activity than the evident effects on the Lorenz and Rikitake systems. As a matter of fact, we are not aware of any direct relation between the positions of the planets and the sunspot number aside from attempts to derive the $11$-year Schwabe cycle from it \cite{stoch_dynamo,hansson}.

We have also explored the effect of perturbation intensity on the coupling in Sect.~\ref{sec:order_param}. There we have gained a broader perspective on the nontrivial effects of a sum of harmonic perturbations on each of the three solar dynamo models. This approach again leads to two forms of testing the validity of solar dynamo models and perturbations in an experimental setting. First, experimental efforts can be made to situate planetary magnetic influences on the Sun along the spectrum of perturbative intensities to see what kind of regime each model proposes. This allows for new ways to evaluate their relative merits. Secondly, exoplanetary observations may help validate and expand our toy models by providing information on more than one value of $\chi$, which may prove invaluable especially for the Lorenz and Rikitake models. The irregular regime in the Duffing-Van der Pol oscillator shown in Fig.~\ref{fig:dvdp_correl} will require a much larger volume of data to explore.

In conclusion, we have seen that adding four harmonic perturbations originates temporal structures consistent with solar observations in a range of nonlinear models of the solar dynamo (based on Lorenz, Rikitake and Duffing Van der Pol systems). These include solar cycles with varying amplitudes and similar duration, as well as hints at long-scale structures in time. The methods of analysis we have explored suggest several new ways to test the validity of the planetary hypothesis stating that solar cycles are critically originated (or dominated) by the magnetic interaction with the planets. The same methods can also be applied to test the existing solar dynamo models against each other. This can be done by introducing physical dimensions into the models to compare perturbative intensities and frequencies with planetary data.

We can also apply the above methods to explore other analytical forms for either the full perturbation function in Eq.~(\ref{eq:perturb_structure}) or the additive components shown in Eq.~(\ref{eq:harmonic_comps}). Moreover, as mentioned in \cite{hansson}, it is theoretically possible to test the proposed magnetic influence of stellar dynamos by observing outside of the Solar System.

\section*{ACKNOWLEDGMENTS}
 This work has been financially supported by the Spanish State Research Agency (AEI) and the European Regional Development Fund (ERDF) under Project No. PID2019-105554GB-I00 (MCIN/AEI/10.13039/501100011033).


\end{document}